\documentclass[aps,prd,reprint,11pt, onecolumn, tightenlines, notitlepage, superscriptaddress, nofootinbib, preprintnumbers, floatfix]{revtex4}

\usepackage{amstext}
\usepackage{amssymb}
\usepackage{amsmath}
\usepackage{graphicx}
\usepackage{hyperref}
\usepackage{url}
\usepackage{color}
\usepackage{ulem}
\usepackage[utf8]{inputenc}
\pdfoutput=1
\usepackage{textcomp}

\usepackage{epsfig,amsfonts,mathrsfs,amsmath,amssymb,graphicx,color,slashed,multirow}
\usepackage{amsmath,latexsym,amssymb,graphicx,slashed,hyperref,color,enumerate,url,cancel,gensymb}
\hypersetup{colorlinks,citecolor= nicered,linkcolor= nicered}
\definecolor{nicered}{rgb}{0.7,0.1,0.1}
\definecolor{nicegreen}{rgb}{0.1,0.5,0.1}

\newcommand{\diag}{\operatorname{diag}}
\def\Valencia{Instituto de F\'{i}sica Corpuscular, CSIC-Universitat de Val\`{e}ncia, 46980 Paterna, Spain}

\begin{document}

\title{{\Large Quasi-Dirac neutrino oscillations at DUNE and JUNO}}

\author{G. Anamiati}\email{anamiati@ific.uv.es}
\affiliation{\Valencia}
\author{V. De Romeri}\email{deromeri@ific.uv.es}
\affiliation{\Valencia}
\author{M. Hirsch}\email{mahirsch@ific.uv.es}
\affiliation{\Valencia}
\author{C. A. Ternes}\email{chternes@ific.uv.es}
\affiliation{\Valencia}
\author{M. T{\'o}rtola}\email{mariam@ific.uv.es} 
\affiliation{\Valencia}
\affiliation{Departament de Física Teòrica, Universitat de València, 46100 Burjassot,  Spain}

\begin{abstract}
Quasi-Dirac neutrinos are obtained when the Lagrangian density of a
neutrino mass model contains both Dirac and Majorana mass terms, and
the Majorana terms are sufficiently small.  This type of neutrinos
introduces new mixing angles and mass splittings into the Hamiltonian,
which will modify the standard neutrino oscillation probabilities. In
this paper, we focus on the case where the new mass splittings are too
small to be measured, but new angles and phases are present. We
perform a sensitivity study for this scenario for the upcoming 
  experiments DUNE and JUNO, finding that they will improve current
bounds on the relevant parameters.  Finally, we also explore the
discovery potential of both experiments, assuming that
neutrinos are indeed quasi-Dirac particles.
\end{abstract}

\maketitle
\newpage\tableofcontents

\section{Introduction}
\label{sec:intro}
Since the discovery of solar and atmospheric neutrino oscillations about two decades ago, 
neutrino oscillation experiments have become more and
more sophisticated. Nowadays many of the parameters characterizing the
conversion of neutrino flavors in the standard 3-neutrino picture
are rather well measured~\cite{deSalas:2017kay}. However, this
framework might not be complete and might need to be extended. Several
studies considering global oscillation data have been performed
assuming the existence of new physics beyond the standard sector, see
for example
Refs.~\cite{Gariazzo:2017fdh,Dentler:2018sju,Esteban:2018ppq}. One of
these scenarios which will be considered here is the case of
quasi-Dirac neutrinos~\cite{Anamiati:2017rxw}.

Since neutrino oscillations are blind to the Dirac or Majorana nature
of neutrinos, one needs other types of experiments, for example, those
searching for neutrinoless double beta decay to determine
it~\cite{Avignone:2007fu,Deppisch:2012nb}.  In general, one can say
that the Dirac case, consisting of $n$ neutrinos, is a
limiting case of the more general Majorana scenario, with $2n$ neutrinos.
This limit is performed by putting the Majorana mass terms
in the Lagrangian to zero. Quasi-Dirac neutrinos arise from the
presence of both Majorana and Dirac mass terms in the Lagrangian
simultaneously, where the Majorana terms are small, but not exactly
zero. As we will show, the departure from Diracness --- i.e. non-zero Majorana
mass terms --- leads to the presence of new mixing angles and new mass
splittings, which will affect  neutrino oscillation probabilities.

Along this work, we will use  ``quasi-Dirac neutrinos" to refer to active-sterile
neutrino pairs~\cite{Valle:1982yw}. In order to distinguish this scenario from the one
with active-active pairs, we denote  the latter ones as pseudo-Dirac
neutrinos~\cite{Wolfenstein:1981kw}.  Many aspects of
pseudo-Dirac neutrinos have been studied in the literature, 
see for example
Refs.~\cite{Petcov:1982ya,Doi:1983wu,Bilenky:1983wt,Bilenky:1987ty,Giunti:1992hk,Dutta:1994wz,Joshipura:2000ts,Nir:2000xn}. 
Note, however, that models with pseudo-Dirac neutrinos do not fit oscillation data
anymore~\cite{Brahmachari:2001rn,Frampton:2001eu,He:2003ih}.  In
the context of quasi-Dirac neutrinos, many papers appeared in the literature proposing
explanations for the solar and atmospheric neutrino
problems~\cite{Geiser:1998mr,Krolikowski:1999an,Balaji:2001fi}, as well as 
consistent descriptions of  standard and short baseline neutrino
oscillations~\cite{Ma:1995gf,Goswami:2001zg}. Several
papers derived limits on quasi-Dirac neutrino properties from
different data sets~\cite{Cirelli:2004cz,deGouvea:2009fp}, while
others  discussed them in the context of neutrino
telescopes~\cite{Beacom:2003eu,Esmaili:2009fk,Esmaili:2012ac,Joshipura:2013yba}.

From a theoretical point of view, there are several options on how
quasi-Dirac neutrinos can be created. They can be produced, for instance, in models
with a singular seesaw~\cite{Stephenson:2004wv,McDonald:2004qx},
double seesaw~\cite{Ahn:2016hhq} or Dirac-seesaw~\cite{Chang:1999pb}
mechanisms. Another possibility is to obtain them from extended gauge
groups~\cite{Sanchez:2001ua,Fonseca:2016xsy} or even in super-gravity
theories~\cite{Abel:2004tt}.

Because of the presence of new spinors in Dirac neutrino models, there
is some overlap between the study of quasi-Dirac neutrinos and the scenario
with sterile neutrinos. 
Several experimental hints point towards the existence of sterile neutrinos, 
which have been extensively investigated in many experiments.
The possible observation of short baseline
oscillations in some of these
experiments~\cite{Athanassopoulos:1996jb,Aguilar:2001ty,Abdurashitov:2005tb,Laveder:2007zz,Giunti:2006bj,Mention:2011rk,Gariazzo:2018mwd,Dentler:2017tkw}
together with the non-observation of neutrino oscillations in
others~\cite{Adamson:2017zcg,Ahmad:2002jz,Aartsen:2017bap,Adamson:2016jku,Agafonova:2015neo,Abe:2014gda,MINOS:2016viw,Adamson:2017uda,Albert:2018mnz,Abe:2019fyx}
lead to large tensions in the global 3+1
picture~\cite{Gariazzo:2017fdh,Dentler:2018sju,Diaz:2019fwt,Gariazzo:inprep}, which cannot be
reconciled even adding more than one sterile
neutrino~\cite{Giunti:2015mwa}. For a recent review on this topic, we
refer the reader to Refs.~\cite{Giunti:2019aiy,Boser:2019rta}. Even though there is
some theoretical overlap, these results would point towards new mass
splittings at the $\sim1$ eV$^2$ scale. Therefore, this type of
oscillations cannot be explained with quasi-Dirac neutrinos, whose additional mass splittings are constrained to be much below the eV scale.

In this paper, we study the sensitivity of the upcoming Deep Underground Neutrino Experiment (DUNE)~\cite{Abi:2018dnh,Abi:2018alz,Abi:2018rgm} 
and Jiangmen Underground Neutrino Observatory (JUNO)~\cite{An:2015jdp} to quasi-Dirac neutrino oscillations. 
The DUNE experiment, hosted by Fermilab, will exploit the synergy of a very high intense neutrino beam
and two massive argon detectors to carry on a broad research program
in neutrino physics. DUNE will allow to perform tests of the three-neutrino paradigm
with remarkable sensitivity, in particular concerning neutrino oscillation 
parameters~\cite{Ghosh:2014rna,DeRomeri:2016qwo,Srivastava:2018ser}. 
The high intensity of the neutrino beam as well as the high resolution of the near and far 
detectors, which characterize DUNE, will make it a leading experiment also in the
search for new physics.
Hence, besides pursuing a comprehensive study of the neutrino mixing, DUNE will also allow 
to explore new physics scenarios, for instance, via the search for non-standard interactions~\cite{Coloma:2015kiu,deGouvea:2015ndi,Blennow:2016etl} or  sterile neutrinos~\cite{Coloma:2017ptb,Agarwalla:2016xxa}, among others~\cite{Escrihuela:2016ube,Masud:2017bcf,Barenboim:2017ewj,Barenboim:2018lpo}. 
JUNO is a next generation reactor experiment and will be located at 53 km from the Yangjiang (six cores with 2.9 GW$_\text{th}$ thermal power each) and Taishan (four cores with 4.6 GW$_\text{th}$ thermal power each) nuclear power plants. The current Daya Bay complex will also contribute with roughly 3\% to the total antineutrino flux. The JUNO detector will be made of 20 kton of liquid scintillator.  With these powerful  sources and an excellent energy resolution, JUNO will be expecting around $10^5$ inverse beta decay events in total. Huge statistics and the long baseline  (for a reactor experiment) assure a measurement of $\sin^2\theta_{12}$ , $\Delta m_{21}^2$ and $\Delta m_{ee}^2$~\cite{Nunokawa:2005nx,Parke:2016joa} at below 1\% level, which makes it a very complementary experiment to DUNE.\\
Our paper is structured as follows. In Sec.~\ref{sec:QD-osc} we present the theoretical framework for quasi-Dirac neutrinos. The simulation of the DUNE and JUNO experiments is described  in Sec.~\ref{sec:sim}. Next, we discuss our results in Sec.~\ref{sec:discussion} and, finally, we draw our conclusions in Sec.~\ref{sec:conc}.

\section{Quasi-Dirac neutrino oscillations}
\label{sec:QD-osc}

A pair of quasi-Dirac neutrinos is a pair of Majorana neutrinos with a
small mass splitting and a relative CP-sign between the two states.
For the sake of illustration, let us start considering only one
neutrino generation. In this case, in the basis
$\left(\nu,N^{c}\right)$, where $\nu$ and $N^{c}$ are the active and
the sterile neutrinos, respectively, the most general neutrino mass
matrix is
\begin{equation}\label{eq:MassMat}
m_{\nu} = 
\begin{pmatrix}
m_{L} & m_{D}\\
m_{D} & m_{R}
\end{pmatrix}\,.
\end{equation}
Here, $m_L$ and $m_R$ are the terms that violate lepton number, while $m_D$
is the standard Dirac neutrino mass term. In the limit in which $m_L$
and $m_R$ are equal to zero, lepton number is conserved and neutrinos
are Dirac particles. This limiting case is characterized by two
degenerate mass eigenstates
\begin{align}
\nu_{1} = & \frac{1}{\sqrt{2}}\left(\nu+N^{c}\right)\nonumber\,, 
\\
\nu_{2} = & \frac{i}{\sqrt{2}}\left(-\nu+N^{c}\right)\,,
\label{eq:degst}
\end{align}
where the factor $i$ is introduced such that both mass eigenvalues
are positive.  Note that, in this mass eigenstate basis, both  $\nu_1$
and $\nu_2$ are equal mixtures of active and sterile neutrinos.
Small deviations from the limit $m_{L}=m_{R}=0$ then lead to
quasi-Dirac neutrinos. If we define the new variables
$\varepsilon=(m_{L}+m_{R})/(2 m_D)$ and $\theta=(m_{L}-m_{R})/(4
m_D)$, in the limit $\varepsilon,\theta \ll 1$, one can rewrite
Eq.~(\ref{eq:degst}) as
\begin{align}
\nu_{1} & \simeq  \frac{1}{\sqrt{2}}\left[\left(1+\theta\right)\nu+\left(1-\theta\right)N^{c}\right]\,,\nonumber
\\
\nu_{2} & \simeq \frac{i}{\sqrt{2}}\left[\left(-1+\theta\right)\nu+\left(1+\theta\right)N^{c}\right]\,,
\label{eq:eqst}
\end{align}
where the quasi-degenerate pairs are nearly maximally
mixed and $\theta$ is a small
angle describing the departure from maximality.
The masses are given by
\begin{equation}
m_{1,2} \simeq m_{D}\left(1\pm\varepsilon\right)\, .\nonumber
\end{equation}
Quasi-Dirac neutrinos are therefore characterized by new mass
splittings and new mixing angles. 

\begin{figure}[!t]
  \centering
  \includegraphics[scale=0.3]{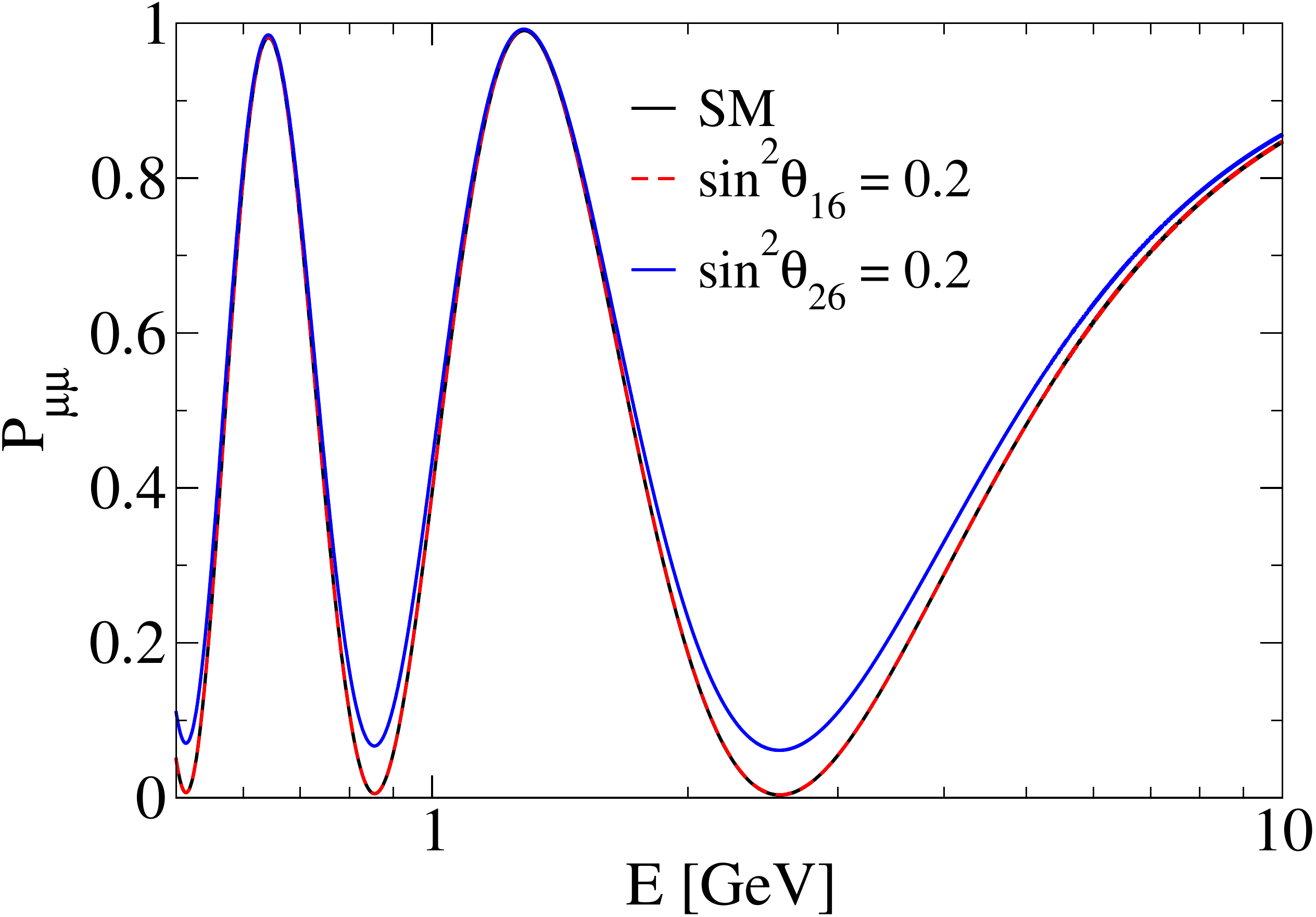}
   \includegraphics[scale=0.3]{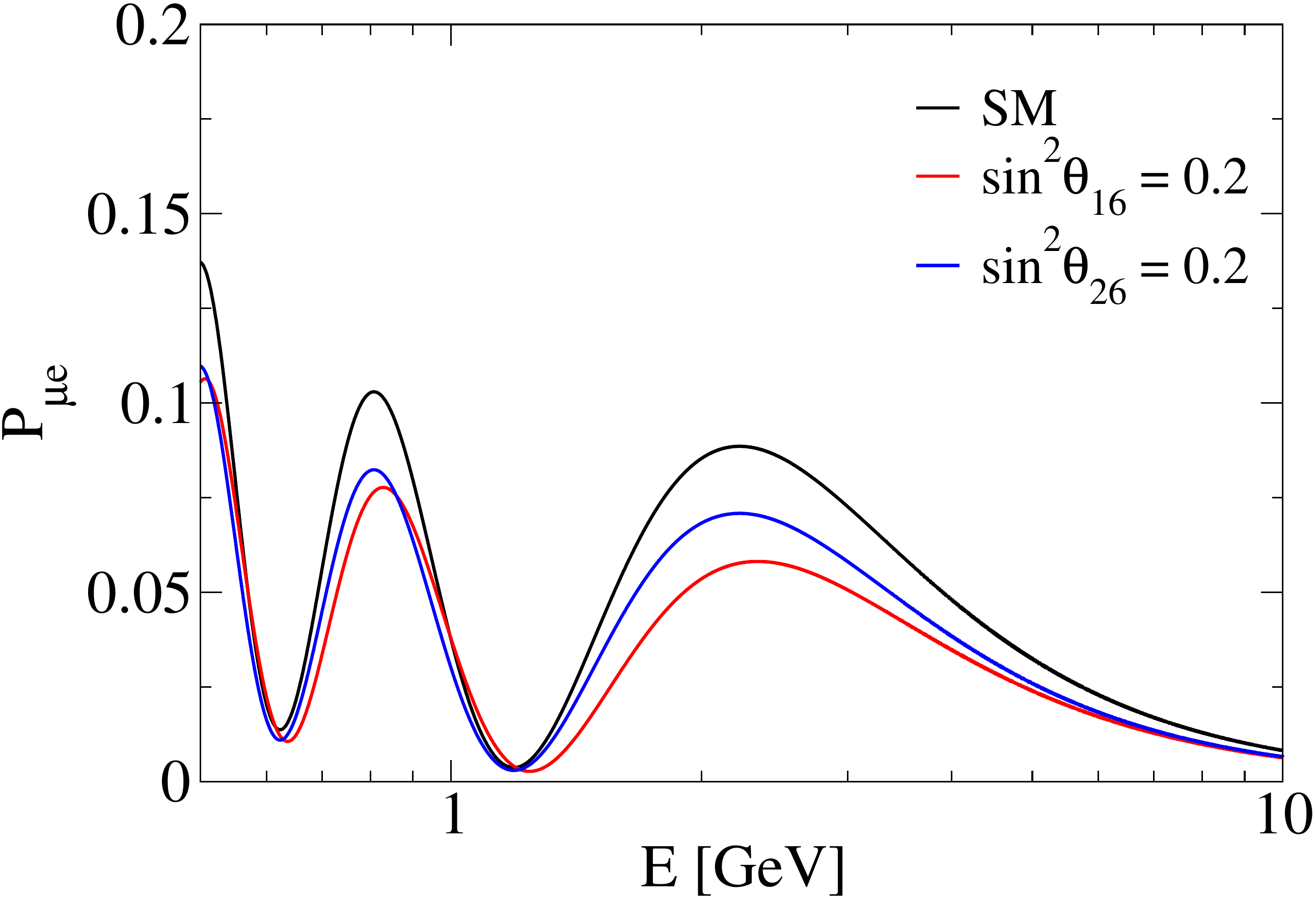}
     \\\includegraphics[scale=0.3]{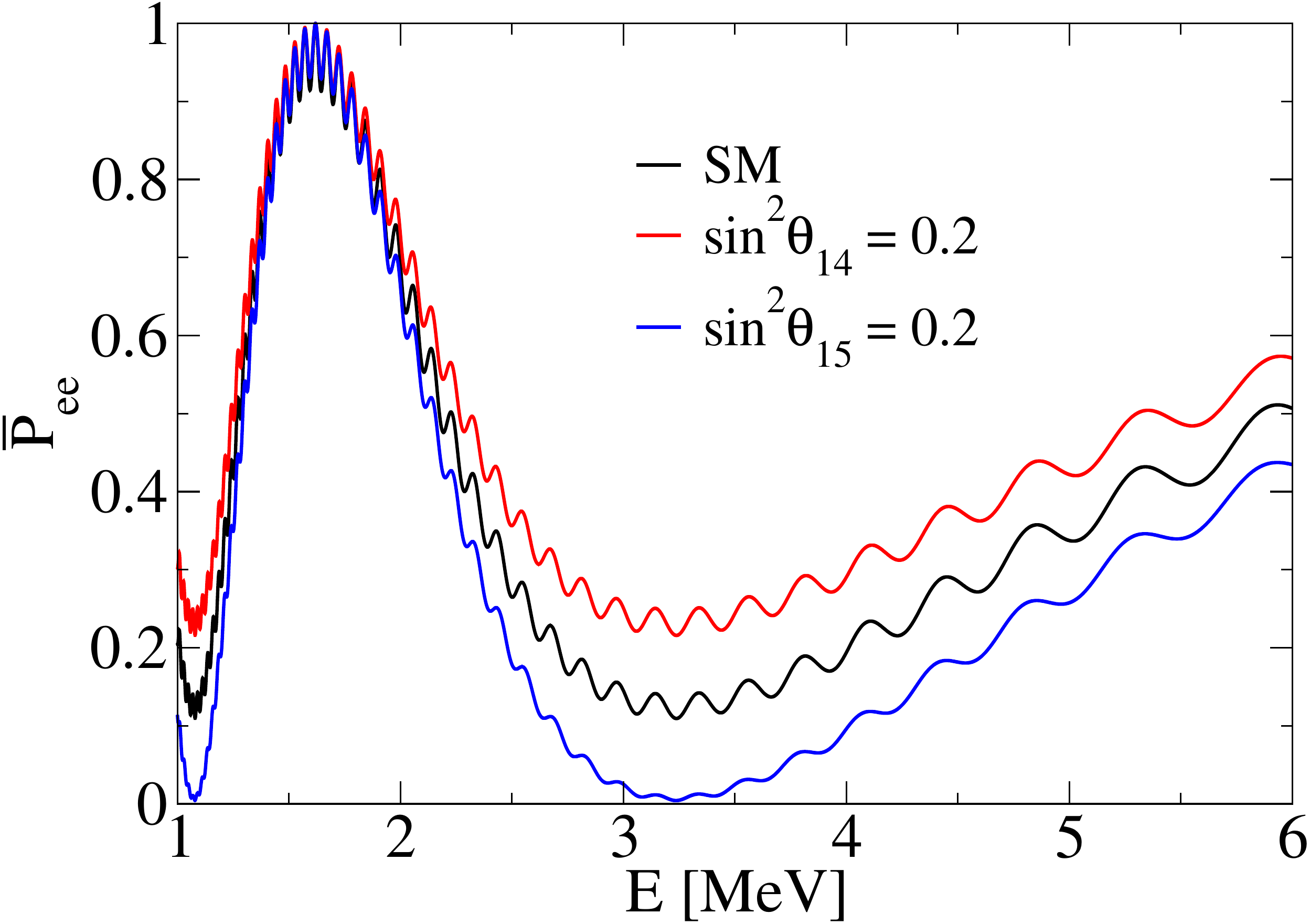}
  \caption{Disappearance (upper left)  and appearance (upper right) probability, $P_{\mu\mu}$ and $P_{\mu e}$, at DUNE as a function of the neutrino energy. Lower panel: antineutrino survival probability, $\overline{P}_{ee}$, in JUNO.    
  In all cases, the black line corresponds to the expected probability in the SM with only three active neutrinos, while the red and blue
    curves are obtained setting a new mixing angle in the quasi-Dirac scenario to the value indicated in the legend. All the other new mixing angles are set to zero.}
  \label{fig:osc}
\end{figure}

Let us now consider the extension of the standard model (SM) with
three sterile neutrinos $N^c$. In the physical mass eigenstate basis,
the charged current SM Lagrangian is modified to
\begin{equation}
\mathcal{L}_{CC}\, =\, -\frac{g}{\sqrt{2}} \, W^-_\mu \,
\sum_{l=1}^{3} \sum_{j=1}^{6} {\bf V}_{lj} \bar \ell_l 
\gamma^\mu P_L \nu_j \, + \, \text{h.c.}\,,
\label{eq:CClagrangian}
\end{equation}
where $P_{L,R} = (1 \mp \gamma_5)/2$ are the chirality projectors, $l
= 1, 2, 3$ denote the flavor of the charged leptons, and $j = 1,
\dots, 6$ the physical neutrino states. The mixing is parameterized by
a rectangular $3 \times 6$ mixing matrix, ${\bf V}_{lj}$
\cite{Schechter:1980gr}. Moreover, the addition of the three sterile
neutrinos allows for the mass term
\begin{equation}
\mathcal{L}_\text{mass} \, = \, 
\frac{1}{2}\, \bar \nu_{\alpha}\, M_{\alpha\beta}\, \nu_{\beta} + \text{h.c.}\,
\label{eq:Lagrangianmass}
\end{equation}
Here, indices $\alpha,\beta = 1, 2, 3$ ($4,5,6$) are for active (sterile)
neutrinos and $M_{\alpha\beta}$ is the generalization of Eq.~(\ref{eq:MassMat})
for three generations.  The full neutrino mass matrix is now
diagonalized by a $6\times6$ unitary matrix, ${\bf \tilde{U}}$.  We
parameterize the neutrino mixing matrix as
\begin{equation}
{\bf \tilde{U}} \left(\theta_{ij},\delta_{ij}\right) = \widehat{R}_{56}\widehat{R}_{46}\widehat{R}_{36}\widehat{R}_{26}\widehat{R}_{16}\widehat{R}_{45}\widehat{R}_{35}\widehat{R}_{25}\widehat{R}_{15}\widehat{R}_{34}\widehat{R}_{24}\widehat{R}_{14}\widehat{R}_{23}\widehat{R}_{13}\widehat{R}_{12}\,,
\label{eq:Umat}
\end{equation}
where $\widehat{R}_{ij}$ are complex rotation matrices which depend on
the mixing angles $\theta_{ij}$ and CP-violating phases $\delta_{ij}$.
The rotation matrices $\widehat{R}_{ij}$ are parameterized  in 
the usual way. For example, for $\widehat{R}_{14}$ we have
\begin{align}
	\widehat{R}_{14}& =
	\begin{pmatrix}
		\cos\theta_{14} & 0 & 0 & e^{-i\delta_{41}}\sin\theta_{14} & 0 & 0 \\
		0 & 1 & 0 & 0 & 0 & 0 \\
		0 & 0 & 1 & 0 & 0 & 0 \\
		-e^{i\delta_{41}}\sin\theta_{14} & 0 & 0 & \cos\theta_{14} & 0 & 0\\
		0 & 0 & 0 & 0 & 1 & 0 \\
		0 & 0 & 0 & 0 & 0 & 1
	\end{pmatrix}	
	\,.\label{eq:elementaryRotation}
\end{align}
Note that the matrix ${\bf \tilde{U}}$ in Eq.~(\ref{eq:Umat}) contains
the mixing among  sterile neutrinos, not observable in neutrino
oscillation experiments. Thus, we will neglect these rotations in the
following.  In the remaining rotations, we have in general 12 angles and 12
phases. However, in our numerical studies we will limit
ourselves to  two phases only, namely $\delta_{13}$ and $\delta_{16}$.
This means that the mixing matrix above can be reduced to
\begin{equation}
	{\bf \tilde{U}}\left(\theta_{ij},\delta_{ij}\right)  = R_{36}R_{26}
\widehat{R}_{16}R_{35}R_{25}R_{15}R_{34}R_{24}R_{14}R_{23}\widehat{R}_{13}R_{12}\,,
	\label{eq:Umatred}
\end{equation}
where $R_{ij}$ denote real rotations.  It proves convenient to
multiply ${\bf \tilde{U}}$ by the following $6\times6$ rotation matrix
(as in Eq.~(16) of Ref.~\cite{Anamiati:2017rxw}):
\begin{equation}
	{\bf U}\left(\theta_{ij},\delta_{ij}\right) \equiv{\bf \tilde{U}}\left(\theta_{ij},\delta_{ij}\right)W,\quad {\rm with}~W=\frac{1}{\sqrt{2}}
	\begin{pmatrix}
		I_3 &  i I_3\\
		I_3 & -i I_3
	\end{pmatrix}\,,
	\label{eq:UtimesI}
\end{equation}
with $I_3$ being the $3\times3$ identity matrix. 
This redefinition allows  to recover trivially the Dirac limit for the mixing 
matrix, by putting to zero all non-standard angles.
The probability of a neutrino oscillating from a flavor
$\alpha$ to a flavor $\beta$ can then be written as
\begin{equation}
	P\left(\nu_{\alpha}\rightarrow\nu_{\beta}\right) 
	= \left|\sum_{j=1}^{6}{\bf U}_{\beta j}{\bf U}_{\alpha j}^{*}
	\exp\left(-\frac{i m_{j}^{2}L}{2E}\right)\right|^{2}\,,
	\label{eq:probabalphabeta}
\end{equation}
where $L$ is the length traveled by the neutrino and $E$ its  energy. 
Therefore, neutrino oscillations are described by the Hamiltonian
\begin{equation}
 \mathcal{H}_0 = \frac{1}{2E}{\bf U}\mathbb{M}^2{\bf U}^\dagger,
\end{equation}
where $\mathbb{M}^2=\diag(0,\Delta m_{21}^2,\Delta
m_{31}^2,\epsilon_1^2, \Delta m_{21}^2+\epsilon_2^2,\Delta
m_{31}^2+\epsilon_3^2)$\footnote{From this expression it is clear why the
  convention chosen in Eq.~(\ref{eq:UtimesI}) is useful. Setting 
  $\epsilon_i^2$ and the non-standard angles to zero,
  $P\left(\nu_{\alpha}\rightarrow\nu_{\beta}\right)$ reduces to the
  standard expression for three generations, despite the fact that we
  sum over six states.} and
   the square of the lightest neutrino mass,  $m_1^2$, has been subtracted from the diagonal elements in the $\mathbb{M}^2$ matrix, 
   as usual. To include matter effects on the neutrino propagation, one should 
add the effective matter potential to the neutrino Hamiltonian above. Quasi-Dirac neutrinos feel the same
potential in the 4-5-6 sector as in the 1-2-3 sector.  Thus,
\begin{equation}
 \mathcal{H} = \frac{1}{2E}\left({\bf U}\mathbb{M}^2{\bf U}^\dagger + \mathbb{A}\right)\,,
\end{equation}
where the potential is now given by $\mathbb{A} =
\diag(V_\text{CC}+V_\text{NC},V_\text{NC},V_\text{NC},V_\text{CC}+V_\text{NC},
V_\text{NC},V_\text{NC})$. 
The charged current potential is given by $V_\text{CC} = 2E\sqrt{2}G_Fn_e$, where $G_F$ is the Fermi
constant and $n_e$ is the electron number density.
The neutral current potential, $V_\text{NC}$, is a  common term to all the diagonal entries and, therefore, it can be removed from the effective Hamiltonian, that will read as follows
\begin{equation}
\mathbb{A} = \diag(V_\text{CC},0,0,V_\text{CC},0,0)\,. 
\label{eq:pot_QD} 
\end{equation}
This Hamiltonian will lead to a different oscillation behaviour
compared to the standard case, as soon as any $\epsilon_i$ or any
non-standard mixing angle is different from zero.  As an example, we
show in Fig.~\ref{fig:osc} (top panels) the oscillation probabilities for the two
channels relevant for DUNE, $\nu_\mu\rightarrow\nu_e$ and
$\nu_\mu\rightarrow\nu_\mu$. The standard oscillation parameters in
these plots are fixed to the ones in Tab.~\ref{tab:oscparam}, taken
from Ref.~\cite{deSalas:2017kay}.  In the left panel we show the
disappearance probability $P_{\mu\mu}$ as a function of the neutrino
energy, turning on one new mixing angle at a time -- which is always
set to $\sin^2\theta_\text{new}=0.2$. The new angle $\theta_{16}$ has
no visible effect on the disappearance probability, while 
$\theta_{26}$ has a visible effect close to the oscillation minima. In
the right panel of Fig.~\ref{fig:osc} we show the appearance
probability $P_{\mu e}$. Here both angles have a visible
impact in the oscillation probability. This is expected from the fact
that the new angles $\theta_{16}$ and $\theta_{26}$ take the role of
the standard angles $\theta_{13}$ and $\theta_{23}$, respectively.
On the other hand, the lower panel of Fig.~\ref{fig:osc}  shows the effect of the new mixing angles
$\theta_{14}$ and $\theta_{15}$ on the survival probability of electron antineutrinos at JUNO. 
In this case, the two mixing angles have opposite effects.
Note that the survival probability shown
here does not include the experimental energy
resolution, which is included in our simulation of JUNO in Sec.~\ref{sec:sim}.

\begin{table}[!t]
\centering
  \catcode`?=\active \def?{\hphantom{0}}
   \begin{tabular}{|c|c|}
    \hline
    Parameter & Value
    \\
    \hline
    $\Delta m^2_{21}$       & $7.5\times 10^{-5}$ eV$^2$\\  
    $\Delta m^2_{31}$       &  $2.50\times 10^{-3}$ eV$^2$\\
    $\sin^2\theta_{12}$     & 0.32\\ 
    $\sin^2\theta_{23}$     & 0.547\\
    $\sin^2\theta_{13}$     & 0.0216\\
    $\delta$                & 1.5$\pi$\\
    \hline
    \end{tabular}
    \caption{Standard neutrino oscillation parameters used in the
      analysis, taken from Ref.~\cite{deSalas:2017kay}.}
    \label{tab:oscparam} 
\end{table}

Fig.~\ref{fig:osc} is meant for illustration purposes only: As we
shall see later in Sec.~\ref{sec:discussion}, standard and
non-standard angles are highly correlated in the quasi-Dirac neutrino
scenario and one can obtain perfect degeneracies among certain
parameters. That is, even very different combinations of angles can
lead to similar oscillation probabilities, which makes the establishment of 
limits on quasi-Dirac angles particularly difficult experimentally. 
For this reason, Ref.~\cite{Anamiati:2017rxw} introduced a 
particular set of variables, $X_i$, which are parameterization 
independent combinations of entries in the neutrino mixing 
matrix ${\bf U}$. Not considering transitions to $\nu_\tau$,
due to the scarcity of $\nu_\tau$ appearance data, 
one can show that only seven independent 
combinations of neutrino mixing angles enter the oscillation 
probabilities. The corresponding $X_i$ are defined as
\begin{align}
 X_1 &= |{\bf U}_{e3}|^2 + |{\bf U}_{e6}|^2\,,\, X_2 = |{\bf U}_{e2}|^2 + |{\bf U}_{e5}|^2\,,\nonumber
 \\
 X_3 &= |{\bf U}_{\mu 3}|^2 + |{\bf U}_{\mu 6}|^2\,,\, X_4 = |{\bf U}_{\mu 2}|^2 + |{\bf U}_{\mu 5}|^2\,,\nonumber
 \\
 X_5 &= |{\bf U}_{e3}{\bf U}^*_{\mu 3} + {\bf U}_{e6}{\bf U}^*_{\mu 6}|^2\,,\,  X_6 = |{\bf U}_{e2}{\bf U}^*_{\mu 2} + {\bf U}_{e5}{\bf U}^*_{\mu 5}|^2\,,\nonumber
 \\
 X_7 &= ({\bf U}_{e3}{\bf U}^*_{\mu 3} + {\bf U}_{e6}{\bf U}^*_{\mu 6})\,({\bf U}_{e2}{\bf U}^*_{\mu 2} + {\bf U}_{e5}{\bf U}^*_{\mu 5}),
 \label{eq:Xdef}
\end{align}
where ${\bf U}$ is the full mixing matrix defined in Eq.~(\ref{eq:UtimesI}). Note that $|X_7|^2=X_5X_6$, i.e.
 only the phase in $X_7$ is a free parameter.
The oscillation probabilities in vacuum can be written in terms of the $X_i$ as~\cite{Anamiati:2017rxw}
\begin{align}
	P\left(\nu_{e}\rightarrow\nu_{e}\right) & =1+\left(1-X_{1}-X_{2}\right)X_{2}\mathcal{A}_{21}+\left(1-X_{1}-X_{2}\right)X_{1}\mathcal{A}_{31}+X_{1}X_{2}\mathcal{A}_{32}\,, \label{eq:Pee_quasiDirac}\\ \label{eq:Pmumu_quasiDirac}
	P\left(\nu_{\mu}\rightarrow\nu_{\mu}\right) & =1+\left(1-X_{3}-X_{4}\right)X_{4}\mathcal{A}_{21}+\left(1-X_{3}-X_{4}\right)X_{3}\mathcal{A}_{31}+X_{3}X_{4}\mathcal{A}_{32}\,,\\
	P\left(\nu_{e}\rightarrow\nu_{\mu}\right) & =-\left(X_{6}+\textrm{Re}X_{7}\right)\mathcal{A}_{21}-\left(X_{5}+\textrm{Re}X_{7}\right)\mathcal{A}_{31}+\textrm{Re}X_{7}\mathcal{A}_{32}+\textrm{Im}X_{7}\left(\mathcal{B}_{21}-\mathcal{B}_{31}+\mathcal{B}_{32}\right)\,,\label{eq:Pemu_quasiDirac}
\end{align}
where $\mathcal{A}_{ij}\equiv-4\sin^{2}\left[\left(m_{i}^{2}-m_{j}^{2}\right)L/\left(4E\right)\right]$
and $\mathcal{B}_{ij}\equiv2\sin\left[\left(m_{i}^{2}-m_{j}^{2}\right)L/\left(2E\right)\right]$.

The usefulness of defining these $X_i$ lies in
the fact that, for a three-generation Dirac scenario, there are only four
independent parameters entering these seven quantities: the three standard
mixing angles and the phase $\delta_{13}$.  Thus, in the Dirac limit, one can 
find three relations among the seven $X_i$:
\begin{gather}\nonumber
	X_{5}=X_{1}X_{3}\, ,\hskip10mm \,X_{6}=X_{2}X_{4}, 
\\
\textrm{Re}\left(X_{7}\right)=\frac{1}{2}\left(1-X_{1}-X_{2}-X_{3}-X_{4}+X_{1}X_{4}+X_{2}X_{3}\right)\,.\label{eq:XsDiracnessConditions}
\end{gather}
Eq.~(\ref{eq:XsDiracnessConditions}) allows to formulate quantitative
tests of ``quasi-Diracness''. We will come back to this in
Sec.~\ref{sec:discussion}. Here we note that, although 
seven $X_i$ are defined here, DUNE will not be sensitive to $X_1$
  and $X_2$, since they depend on the solar parameters. However, JUNO
  (and Daya Bay, which we will include as a prior in our analysis) will provide stringent constraints on $X_1$ and $X_2$,
  see section~\ref{sec:discussion}.  On the contrary, DUNE will be
  able to put severe restrictions on $X_3$ and $X_5$ and some
  improvements on the remaining parameters $X_4$, $X_6$ and $X_7$, as
  we will show below.

\begin{figure}
  \centering
  \includegraphics[scale=0.35]{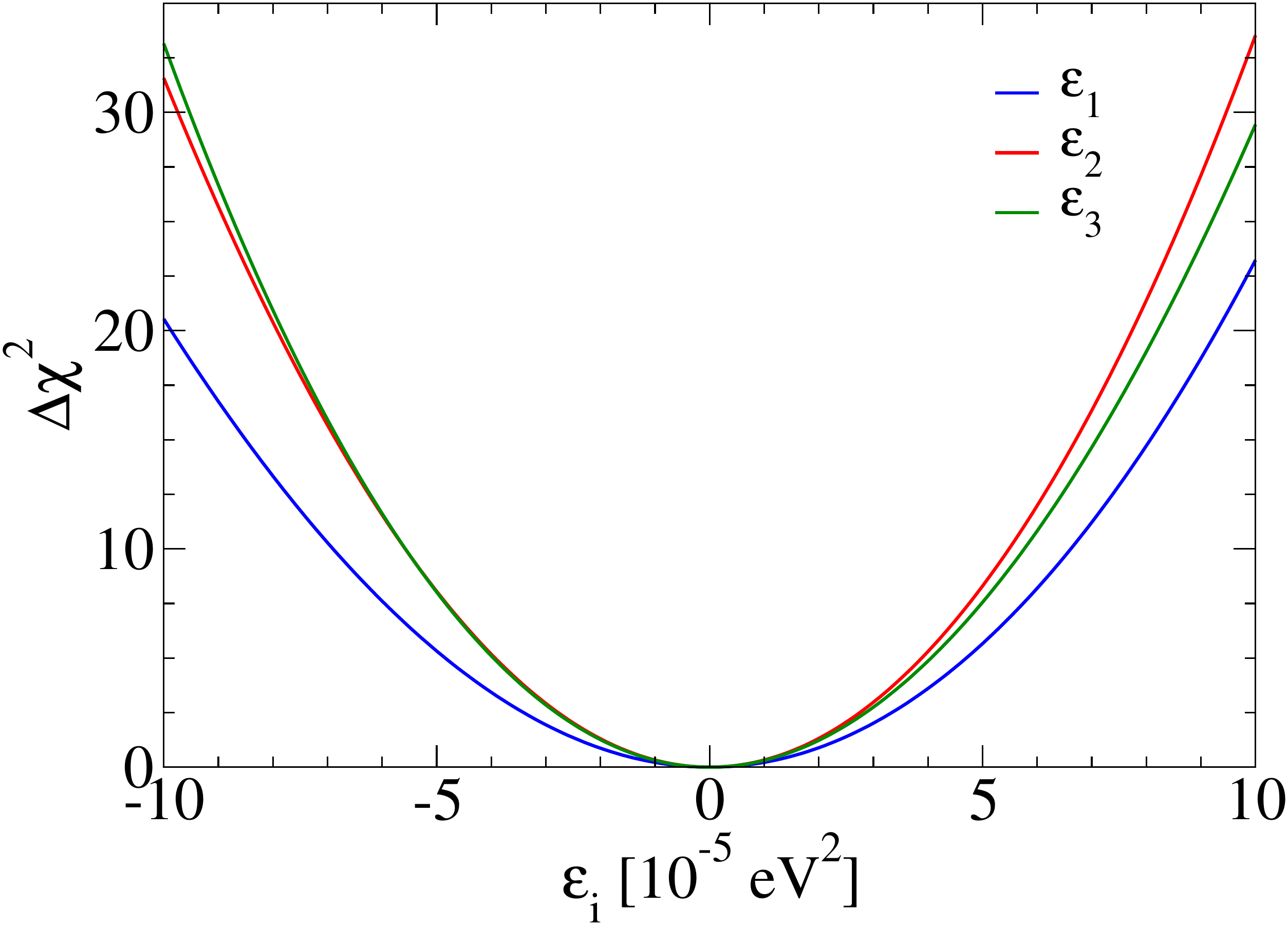}
    \caption{$\chi^2$ profiles for the new mass splittings $\epsilon_i$. New angles are fixed to zero.}
  \label{fig:profiles_eps}
\end{figure}

Beyond the new mixing angles, we show  DUNE's sensitivity to the new mass splittings $\epsilon_i$ in Fig.~\ref{fig:profiles_eps}. These results have been obtained by varying only one of the new mass splittings at a time and fixing the new angles to zero. 
In comparison with previous results derived in Ref.~\cite{Anamiati:2017rxw}, one can see that DUNE will not be competitive with other current experiments, which give bounds on $\epsilon_1$ and $\epsilon_2$ several orders of magnitude stronger than the ones shown in Fig.~\ref{fig:profiles_eps}. The only comparable bound is the one for $\epsilon_3$. Note, however, that in Ref.~\cite{Anamiati:2017rxw} the Authors marginalized over some of the oscillation parameters, while we kept all of them fixed. Marginalizing over additional parameters would result in weaker bounds, also for $\epsilon_3$. Therefore, given the poor sensitivity of DUNE to the  new splittings $\epsilon_i$, we will set them to very small values in our analysis.
The sensitivity of JUNO to the mass splittings $\epsilon_i$ has been discussed in Ref.~\cite{Anamiati:2017rxw}. We can infer from Tab.~1 in~\cite{Anamiati:2017rxw} that JUNO will neither be able to improve the current bounds on any of the new mass splittings.

\section{Simulation of experiments}
\label{sec:sim}

The Deep Underground Neutrino
Experiment~\cite{Abi:2018dnh,Abi:2018alz,Abi:2018rgm} is one of the
next generation long-baseline accelerator experiments. It will consist
of two detectors exposed to a megawatt-scale neutrino beam produced at
Fermilab. This beam will consist of (nearly) only muon neutrinos. The
near detector will be placed approximately 600 meters away from the
source of the beam. The second (far) detector, divided into four
modules, each using 10~kton of argon as detection material, will be
installed 1300 kilometres away deep underground at the Sanford
Underground Research Facility in South Dakota.

To simulate the neutrino signal in DUNE we use the GLoBES package \cite{Huber:2004ka,Huber:2007ji} with the  configuration file provided by the DUNE collaboration \cite{Alion:2016uaj}. We assume DUNE to run 3.5 years in neutrino mode and other 3.5 years in antineutrino mode. Considering an 80 GeV beam with 1.07 MW beam power, this corresponds to  an exposure of 300 kton-MW-years. 
In this configuration, DUNE will be using $1.47\times 10^{21}$ protons on target (POT) per year. Our analysis includes disappearance and appearance channels, simulating both signals and backgrounds. The simulated backgrounds include contamination of antineutrinos (neutrinos) in the neutrino (antineutrino) mode, and also misinterpretation of flavors. 

To include quasi-Dirac neutrino oscillations in our simulation of DUNE, we use the GLoBES extension {\tt snu.c}~\cite{Kopp:2006wp, Kopp:2007ne}. This extension was originally made to include non-standard neutrino interactions and sterile neutrinos in GLoBES simulations. For this analysis, we have modified the definition of the neutrino oscillation probability function inside 
  {\tt snu.c} by adding the additional rotation matrix of Eq.~(\ref{eq:UtimesI}) and the matter potential of Eq.~(\ref{eq:pot_QD}).

For the statistical analysis, we create a fake DUNE data sample using
the standard oscillation parameters from
Tab.~\ref{tab:oscparam}. Next, we try to reconstruct the simulated
data varying the mixing angles $\theta_{13}$, $\theta_{23}$, $\theta_{16}$
and $\theta_{26}$ (most relevant for DUNE) and the two CP-violating phases $\delta_{13}$ and
$\delta_{16}$. The remaining new mixing angles are fixed to zero and the new mass
splittings $\epsilon_i^2$ are fixed to very small values. 
Note as well that, since DUNE has no
sensitivity to the solar parameters, these are fixed at their best fit values, in Tab.~\ref{tab:oscparam}. On the other hand, given that we are mostly interested in correlations between the standard and new mixing angles, we have also kept $\Delta m_{31}^2$ fixed to its best fit value. Currently, there is a  preference for normal mass
ordering slightly above $3\sigma$~\cite{deSalas:2017kay,Gariazzo:2018pei,deSalas:2018bym}, so we will not
consider negative values of $\Delta m_{31}^2$ here.  We use GLoBES to calculate the
event numbers for a given set of oscillation parameters $p$ and then we
calculate the $\chi^2$ value for this set using the following expression
\begin{equation}
 \chi_{\text{DUNE}}^2(p)=\min_{\vec{\alpha}}
 \sum_\text{channels}2\sum_n \left[ N_n(p,\vec{\alpha})- N_n^{\text{dat}} +
 N_n^{\text{dat}} \log \left(\frac{N_n^{\text{dat}}}{N_n(p,\vec{\alpha})}\right)\right] 
 + \sum_i \left(\frac{\alpha_i}{\sigma_i}\right)^2 .
 \label{main-chi2}
\end{equation}
Here, $N_n^\text{dat}$ corresponds to the simulated event number in the
$n$-th bin for the oscillation parameters in Tab.~\ref{tab:oscparam},
$N_n(p,\vec{\alpha})$ is the event number predicted  in the $n$-th bin associated
to the oscillation parameters $p$ and to the nuisance parameters $\alpha_i$, with standard deviations given by $\sigma_i$.
All the nuisance parameters are associated to normalization uncertainties of signal or background events and introduce modifications of the type $N_n \to N_n(1+\alpha_i)$.
The last term in Eq.~(\ref{main-chi2})  penalizes the deviation of the latter parameters from their expectation values, $\alpha_i = 0$.
Finally, the $\chi^2$ sums over disappearance and appearance channels in both neutrino and antineutrino modes. 
The simulation and sensitivity analysis  of the  JUNO reactor experiment are performed following the procedure described in~\cite{Anamiati:2017rxw}. The corresponding $\chi^2$ function, $\chi^2_\text{JUNO}$, is obtained  by allowing the variation of only five parameters which are relevant for JUNO,  namely $\theta_{12}, \theta_{13}, \theta_{14}, \theta_{15}$ and $\theta_{16}$. The solar mass splitting $\Delta m_{21}^2$ is fixed to the best fit value. 
To get the global future sensitivity to the quasi-Dirac scenario, in our analysis we combine the individual sensitivities obtained for DUNE and JUNO. Besides the  two $\chi^2$ functions  discussed above, $\chi^2_\text{DUNE}$ and $\chi^2_\text{JUNO}$, we introduce a penalty function associated to some of the mixing angles under study.
As it was shown in~\cite{Anamiati:2017rxw}, the current reactor experiments cannot univocally measure the reactor angle $\theta_{13}$ in presence of quasi-Dirac neutrinos. 
However, it is still possible to simultaneously constrain several of these angles. If not, Daya Bay would have observed a different signal. This penalty can be obtained from Eq.~(\ref{eq:Pee_quasiDirac}) by imposing
$
\left(1-X_{1}-X_{2}\right)X_{1}+X_{1}X_{2} = \sin^2\theta_{\text{DB}}
$ 
where $\sin^2\theta_{\text{DB}} \approx 0.022$ is the value currently measured by the Daya Bay reactor experiment~\cite{Adey:2018zwh}. Hence, our global $\chi^2$ function can be written as
\begin{equation}
 \chi^2(p)=\chi_{\text{DUNE}}^2(p) + \chi_{\text{JUNO}}^2(p) + \textit{f}_{\text{DB}}(p)\, .
 \label{complete-chi2}
\end{equation}
The penalty function in terms of the relevant mixing angles is given by
\begin{equation}
\textit{f}_{\text{DB}}(p) = \left[ \frac{\left((c_{14} c_{15} c_{16} s_{13})^2 + s_{16}^2 - 1\right) \left((c_{14} c_{15} c_{16} s_{13})^2 + s_{16}^2\right)  - \sin^2\theta_{\text{DB}}}{\sigma_{\text{DB}}  \sin^2\theta_{\text{DB}}}\right]^2\, ,
\label{eq:penalty}
\end{equation}
where $c_{ij} = \cos\theta_{ij}$,  $s_{ij} = \sin\theta_{ij}$ and $\sigma_{\text{DB}}$ is the expected uncertainty in the final measurement of the reactor mixing angle by Daya Bay, set to 3\%. This is a generalization of the standard reactor prior used in several studies on neutrino oscillations. 

\section{Results and discussion}
\label{sec:discussion}

\begin{figure}
  \centering
  \includegraphics[scale=0.5]{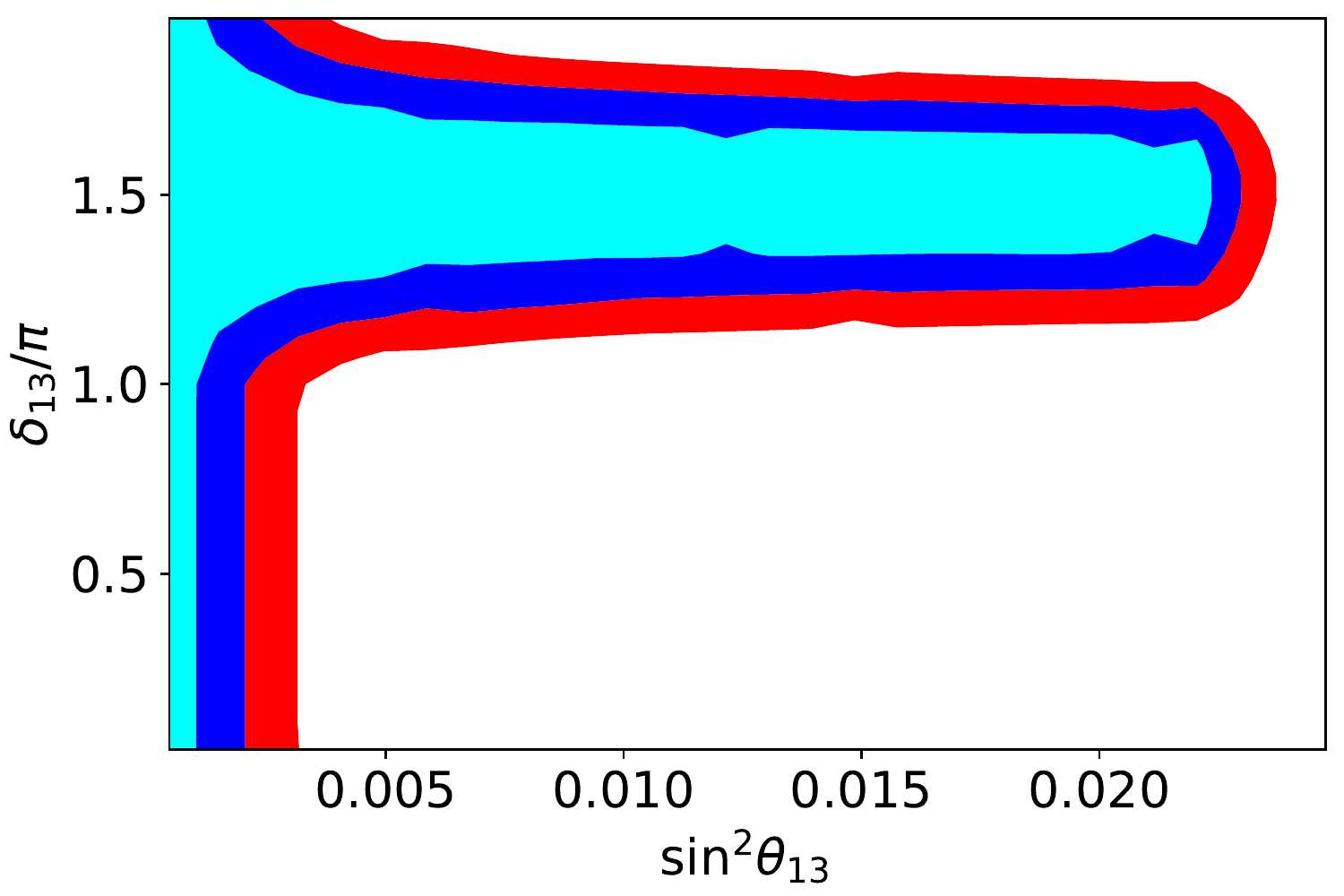}
    \includegraphics[scale=0.5]{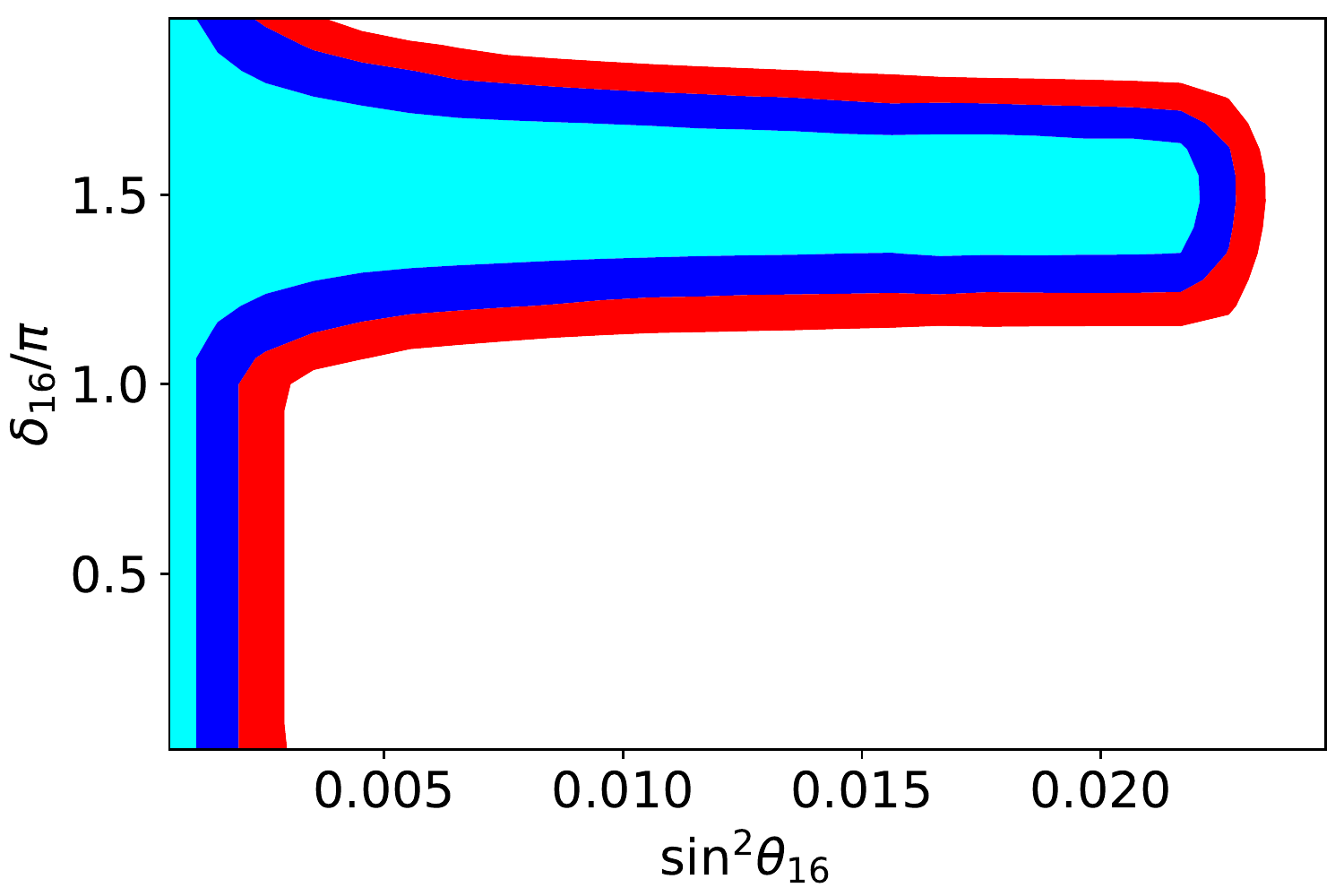}\\
      \includegraphics[scale=0.5]{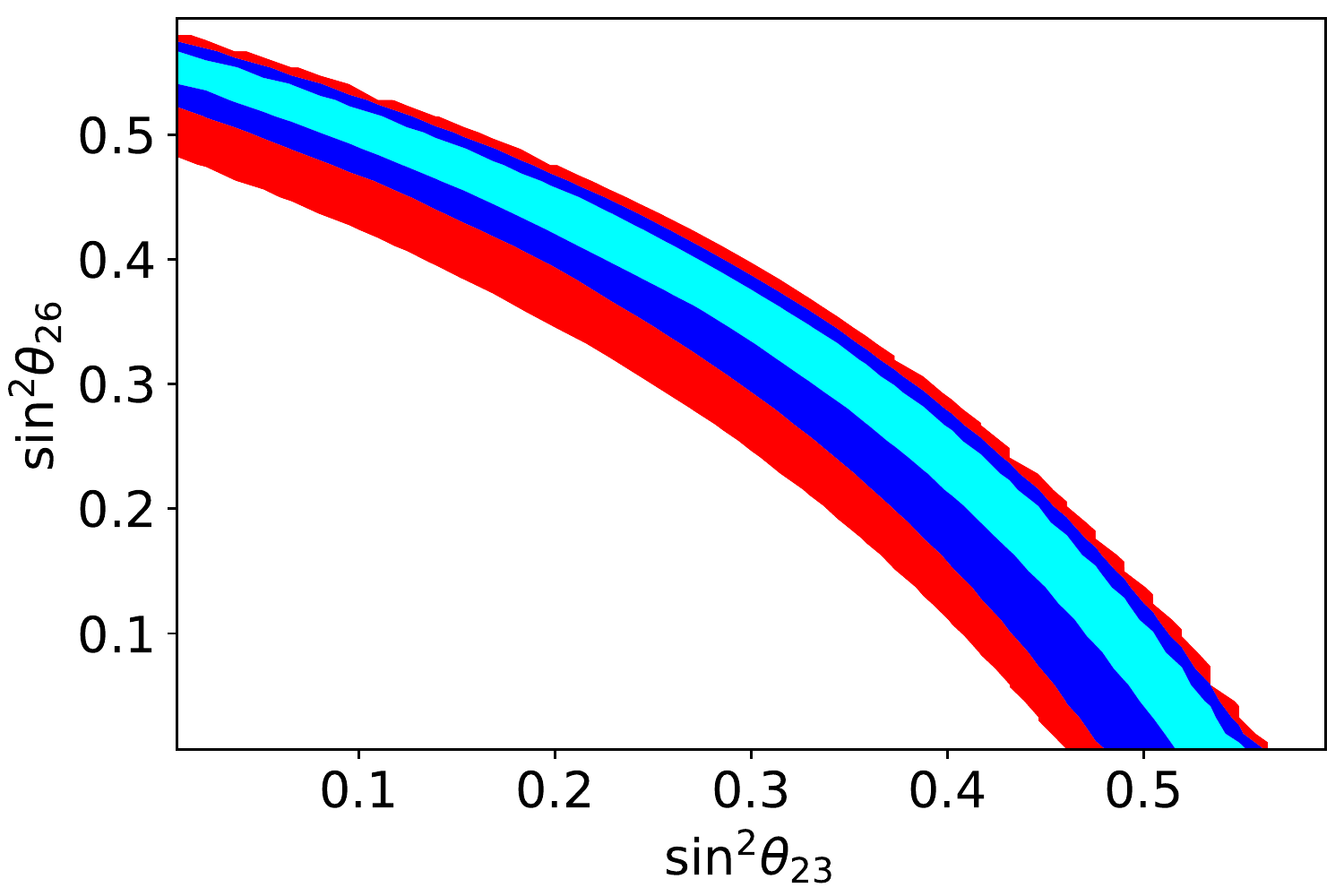}
        \includegraphics[scale=0.5]{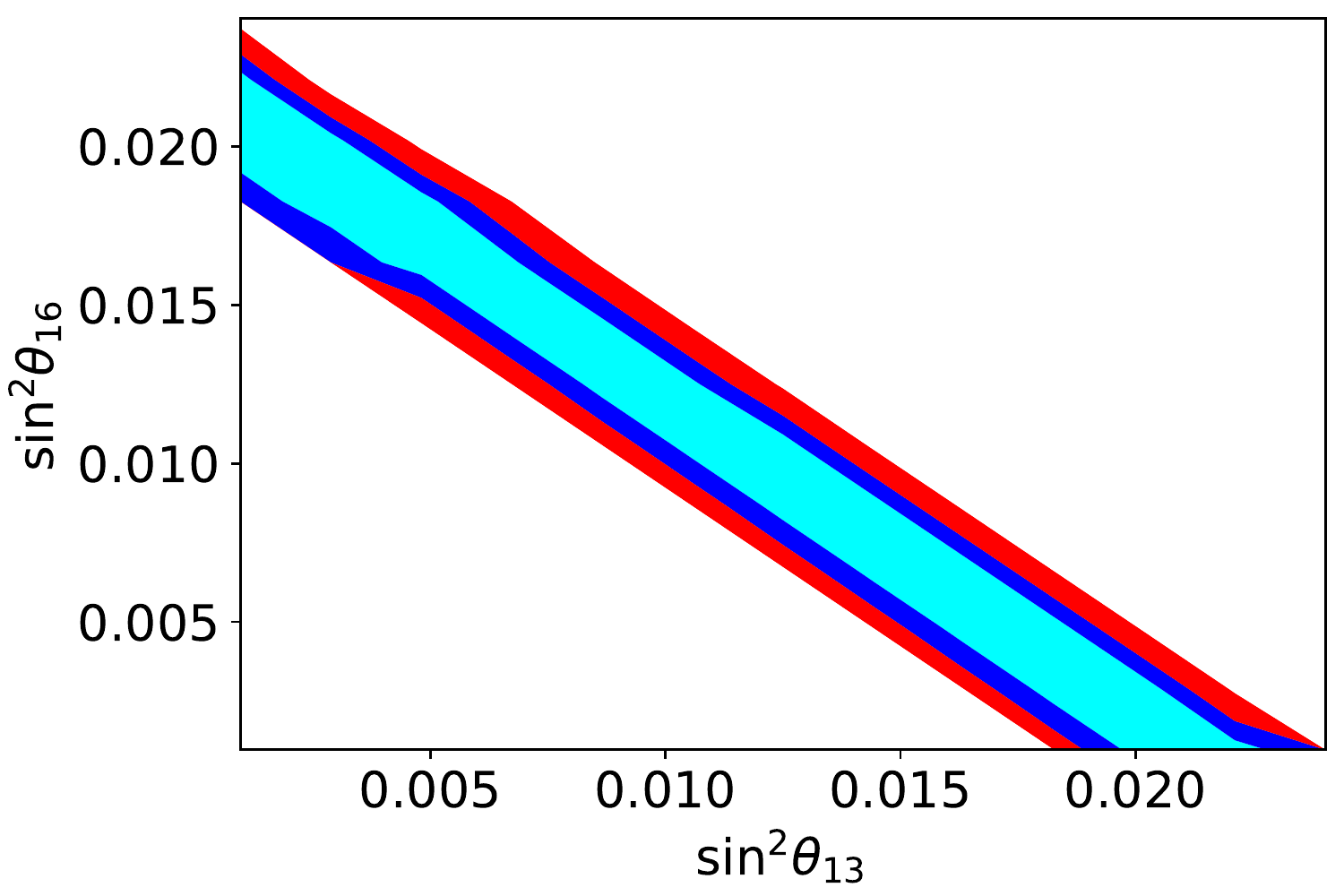}\\
         \includegraphics[scale=0.5]{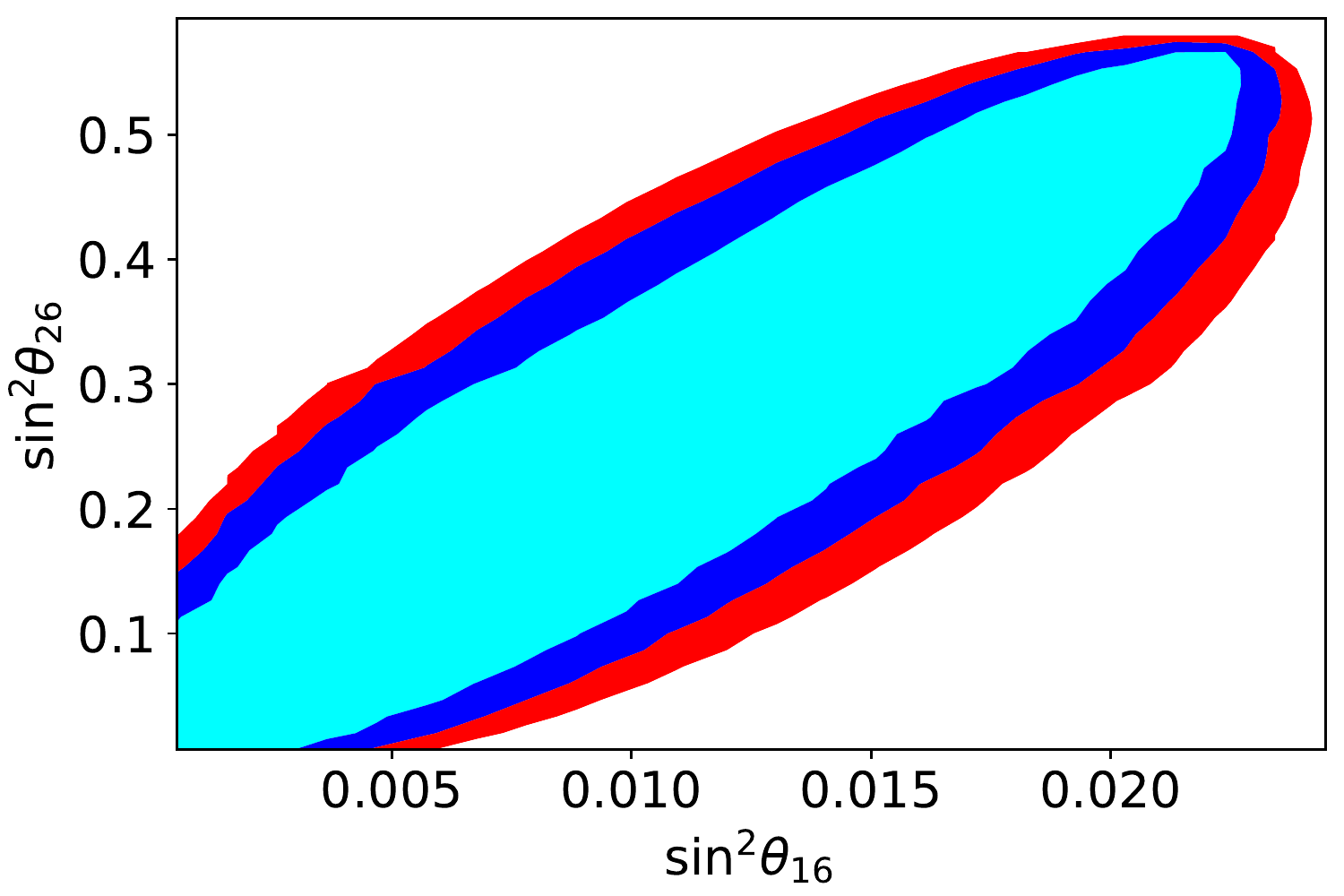}
        \includegraphics[scale=0.5]{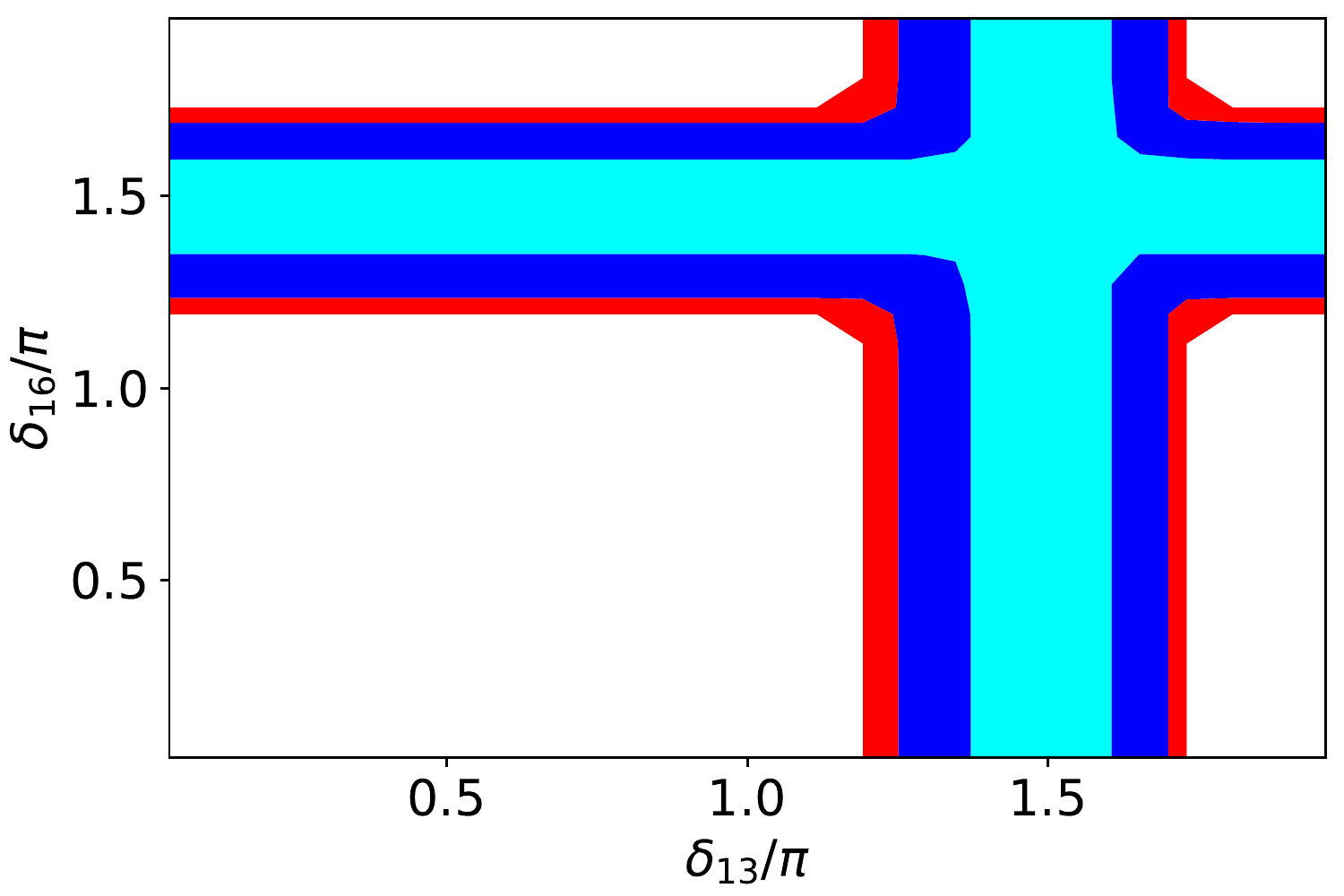}
   \caption{DUNE sensitivity to the oscillation parameters under study. The colored regions shown correspond to 1 (cyan), 2 (blue), 3 (red) $\sigma$ confidence levels for 2 degrees of freedom. In each two-dimensional plot we have marginalized over the other parameters which are not displayed. }
  \label{fig:2dim}
\end{figure}

In this section, we present the results of the statistical analysis performed in this work. Before discussing the results of the combined analysis of DUNE and JUNO, we discuss the results of the two experiments separately. Note, however, that we always add the penalty term in Eq.~(\ref{eq:penalty}) to the $\chi^2$ function obtained from the sensitivity analysis of each experiment. In Fig.~\ref{fig:2dim} we show the two-dimensional allowed regions obtained by scanning over the parameters $\theta_{13}$, $\theta_{23}$, $\theta_{16}$, $\theta_{26}$, $\delta_{13}$ and $\delta_{16}$ in DUNE. The parameters not shown are marginalized over in each panel. The colored regions correspond to the 1 (cyan), 2 (blue), 3 (red) $\sigma$ confidence levels for 2 degrees of freedom.
In the upper panels, we see that the two reactor angles, $\theta_{13}$
and $\theta_{16}$, and their corresponding phases, $\delta_{13}$ and
$\delta_{16}$, behave in a very similar way. In
principle, small values of the phases are allowed, although these
require very small values for the associated mixing angles. From the
right panel of the second row, however, we see that both angles cannot
 be small at the same time: if $\theta_{13}$ is small,
$\theta_{16}$ has to be large and vice versa. In the former case,
$\delta_{13}$ can take any value in the interval $[0,2\pi]$, while
$\delta_{16}$ is rather restricted around its maximal value
$1.5\pi$. This is an interesting point, because the fake data were
created with $\delta_{16} = 0$. The reason behind this is that
  the new angles and phases are correlated to the standard angles, e.g.
  $\theta_{13}$ and $\theta_{16}$ (see the definitions in
  Eq.~(\ref{eq:Xdef})), hence they are interchangeable. Note as well
that all the sensitivity to the reactor angle is lost, since DUNE can
only reproduce the prior~\cite{Anamiati:2017rxw} that we introduced as
an input for our analysis, as explained in Sec.~\ref{sec:sim}.
The interchangeability of the mixing parameters can also be seen from the left panel of the second row in Fig.~\ref{fig:2dim}. There, we see that $\theta_{23}$ and $\theta_{26}$ are also fully correlated: having  a large $\theta_{23}$ and a small $\theta_{26}$ is equivalent to having a small $\theta_{23}$ and a large $\theta_{26}$. The same applies to the CP-violating phases, as can be seen in the right panel of the last row. The left panel of the last row shows that there are correlations also between the atmospheric and reactor angles, which are not 
 present   in the standard case of three-neutrino oscillations anymore, given the very good level of precision achieved in the determination of the mixing angles.
Since $\theta_{16}$ and $\theta_{13}$, as well as  $\theta_{26}$ and  $\theta_{23}$, are equivalent, a similar result is obtained in the two-dimensional plane ($\theta_{13}$, $\theta_{23}$).\\
%


{In Fig.~\ref{fig:2dim-JUNO} we show the result of our simulation of JUNO. In this case, we find that the new angles $\theta_{14}$ and $\theta_{15}$ are highly correlated with the standard solar angle $\theta_{12}$. In particular, one sees that, for $\sin^2\theta_{14} = 0$, all values of $\theta_{12}$ and $\theta_{15}$ lying along the correspondingly labeled line are possible, showing a similar correlation as in the case of $\theta_{13}$ and $\theta_{16}$ or $\theta_{23}$ and $\theta_{26}$. For different values of $\sin^2\theta_{14}$ the correlating line is shifted as indicated in the figure. If we now marginalize over all possible values for $\theta_{14}$, we find that a large region of parameter space is still allowed. It is however important to notice that a point in the ($\sin^2\theta_{12}$, $\sin^2\theta_{15}$) plane always corresponds to one specific value of $\sin^2\theta_{14}$.

\begin{figure}
  \centering
  \includegraphics[scale=0.75]{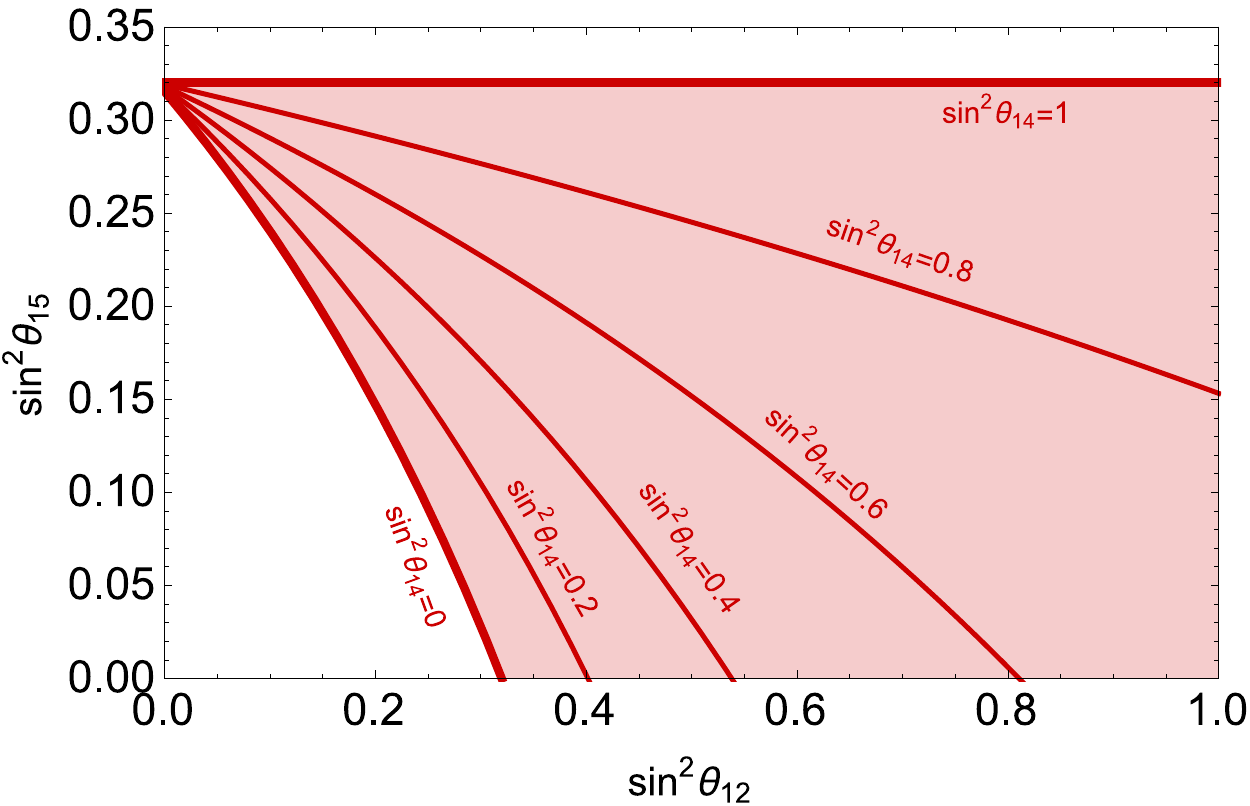}
   \caption{ Sensitivity region in the ($\sin^2\theta_{12}$, $\sin^2\theta_{15}$) plane  for the JUNO experiment. The shaded region corresponds to the 1$\sigma$ allowed region once marginalized over $\sin^2\theta_{14}$. See the text for more details.}
  \label{fig:2dim-JUNO}
\end{figure}


Now let us discuss the results of the combined
  analysis of DUNE and JUNO in terms of the variables $X_i$ introduced
  in Eq.~(\ref{eq:Xdef}). Compared to Ref.~\cite{Anamiati:2017rxw},
  where most of the parameter space was allowed by current neutrino
  oscillation data, here we find that DUNE and JUNO will be able to
  strongly constrain some of the $X_i$ parameters. This is
illustrated in Fig.~\ref{fig:chi2profiles_X}, where we show the
$\Delta\chi^2$ profiles for the $X_i$ variables. In the left panel, we
see how precise DUNE and JUNO could measure some of these
quantities.  Notably, $X_1, X_2$ and $X_3$ can be measured with a
  precision below $\%$.The sensitivity to $X_4$, $X_5, X_6$
  and $X_7$ will also be improved with respect to the current results
  obtained in \cite{Anamiati:2017rxw}, although not as dramatically as
  for the previous three parameters. Note, however, that DUNE
  will not be able to set a lower limit on $X_4$, $X_6$ and $X_7$,
  which are allowed to be zero in our combined fits. In the  case of $X_4$,
  this can be traced to the fact that DUNE does not have the
  resolution to demonstrate that there are three independent
  oscillation frequencies contributing to
  $P\left(\nu_{\mu}\rightarrow\nu_{\mu}\right)$ (see
  Eq.~(\ref{eq:Pmumu_quasiDirac})).  An upper limit on $X_4$ can
  instead be obtained from the unitarity relation $X_{3}+X_{4} < 1$.
  Similar comments apply to $X_6$ and $X_7$.

In the right panel of Fig.~\ref{fig:chi2profiles_X}, we construct a quantity to test directly the Diracness of neutrino oscillations. This quantity is obtained by assuming that neutrinos are Dirac particles in Eq.~(\ref{eq:Xdef}), see Ref.~\cite{Anamiati:2017rxw} for more details. In this case one can derive that
\begin{equation}
 1-\frac{X_5}{X_1X_3} = 0\,.
 \label{eq:Diracness}
\end{equation}
Any deviation from zero in this expression  would be an indication for quasi-Dirac neutrinos. Since we created our fake data assuming neutrinos to be Dirac particles, our best-fit point is automatically located at zero. However, DUNE could restrict the allowed deviation considerably, as shown in the plot.

Finally, to further investigate the discrimination power of the experiments to the quasi-Dirac scenario, we have created another fake data set using a quasi-Dirac point as an input. For the particular point we have chosen, we expect 
\begin{equation}
 1-\frac{X_5}{X_1X_3} = 0.5\,.
 \label{eq:Diracness2}
\end{equation}
Our choice falls inside the 1$\sigma$ contours of Fig.~\ref{fig:2dim} and it corresponds to $\sin^2\theta_{23} = 0.30$, $\sin^2\theta_{26} = 0.37$, $\sin^2\theta_{13} = 0.0108$, $\sin^2\theta_{16} = 0.0108$. The CP-violating phases are assumed as in the first analysis. The result for this simulation is shown in Fig.~\ref{fig:chi2profiles_XQD}. In the left panel one can see how most of the $X_i$ are mostly unaffected by the selected input point, while there is a visible difference in the profiles corresponding to $X_3$ and $X_5$. Nevertheless, the most visible effect appears in the right panel of the figure. There, we see that the Dirac-point --- with $1-\frac{X_5}{X_1X_3} = 0$ --- could be completely excluded in this scenario. This is an important result, because it means that DUNE and JUNO would be able to distinguish standard three-neutrino oscillations from quasi-Dirac oscillations. Note, however, that this statement is true for our benchmark point. If the true value lies very close to Diracness, it would be  more difficult to discriminate between the two scenarios.

\begin{figure}
  \centering
  \includegraphics[scale=0.64]{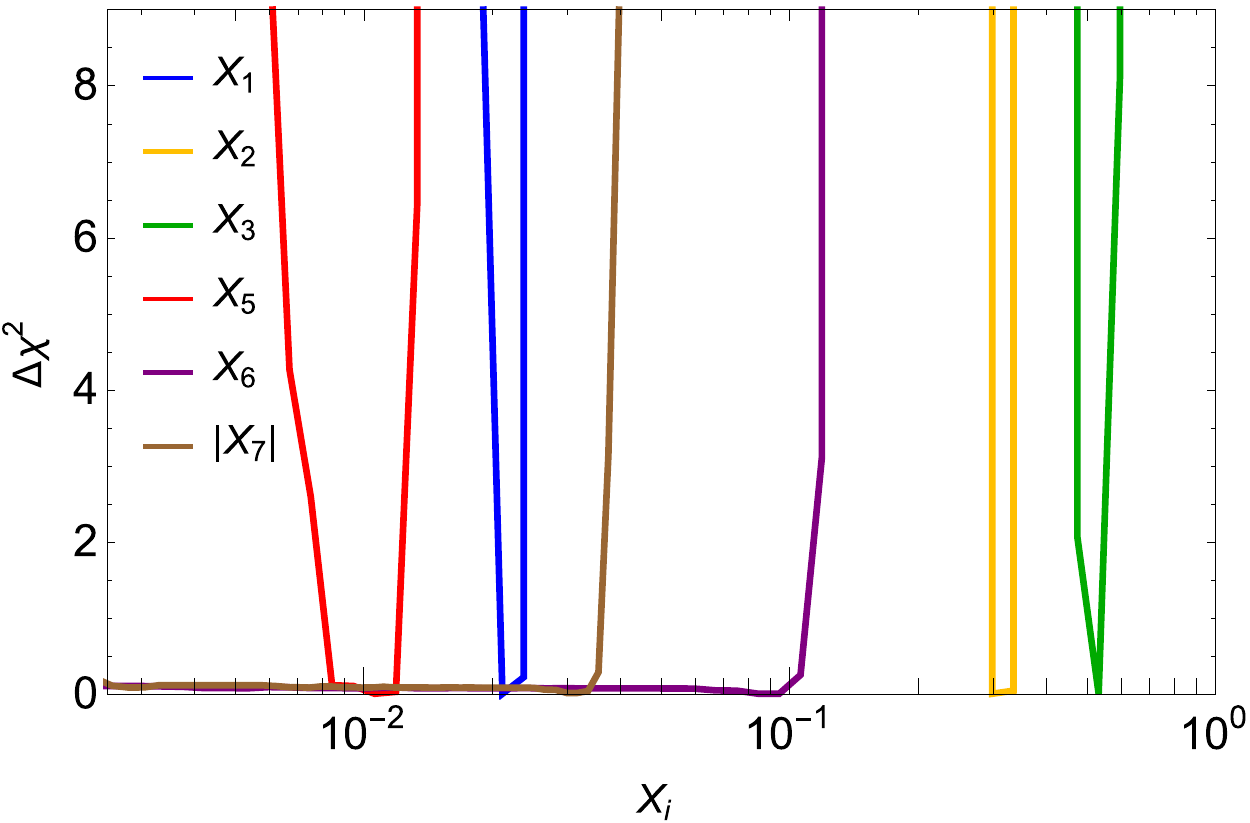}
  \includegraphics[scale=0.64]{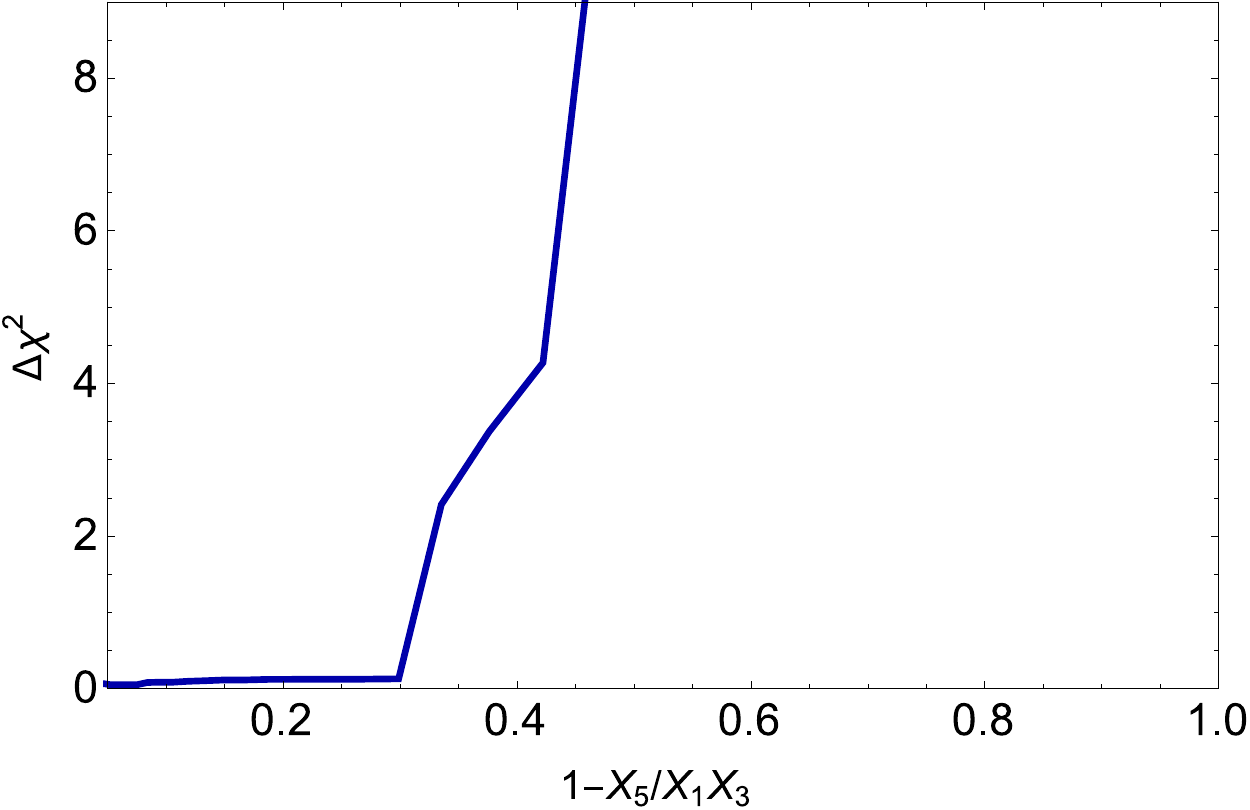}
   \caption{Left panel: $\chi^2$ profiles for the $X_i$ variables. Right panel:  Diracness test for a Dirac input point.}
  \label{fig:chi2profiles_X}
\end{figure}

\begin{figure}
  \centering
  \includegraphics[scale=0.64]{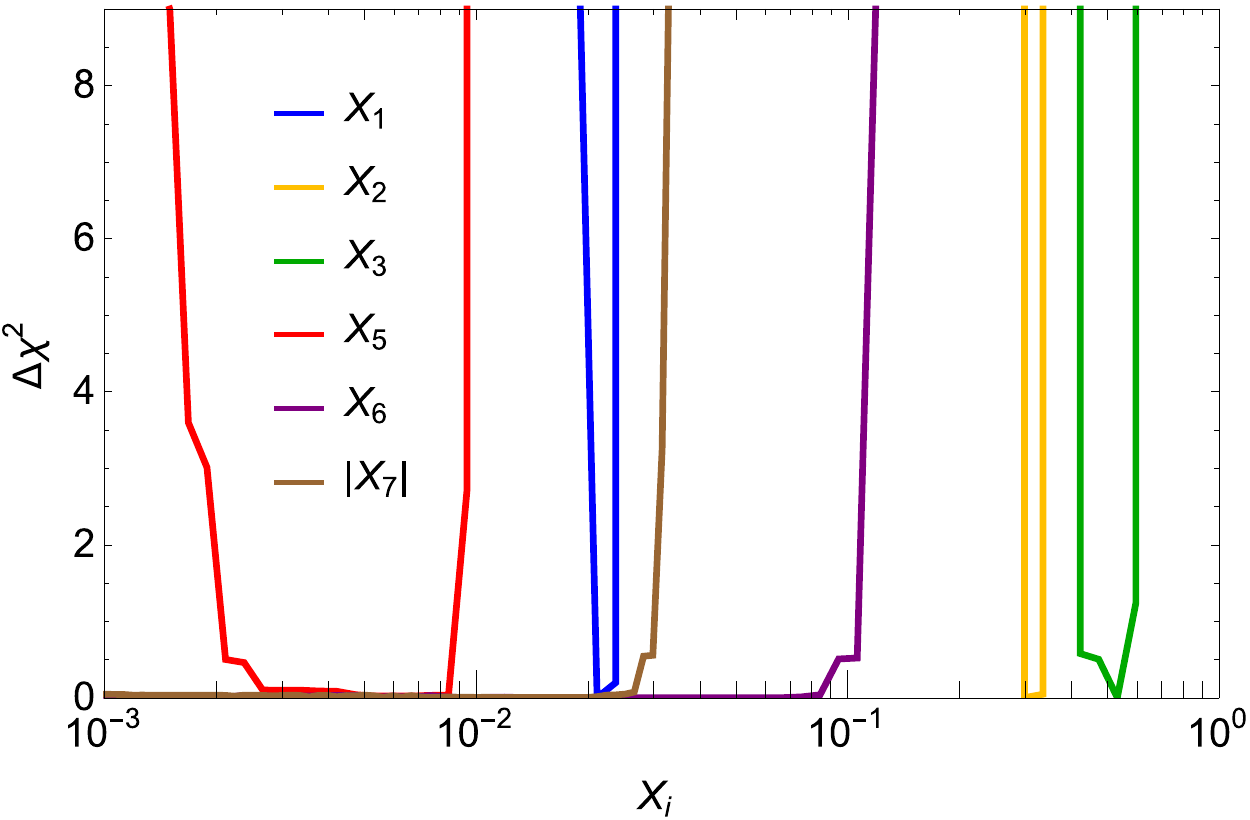}
  \includegraphics[scale=0.64]{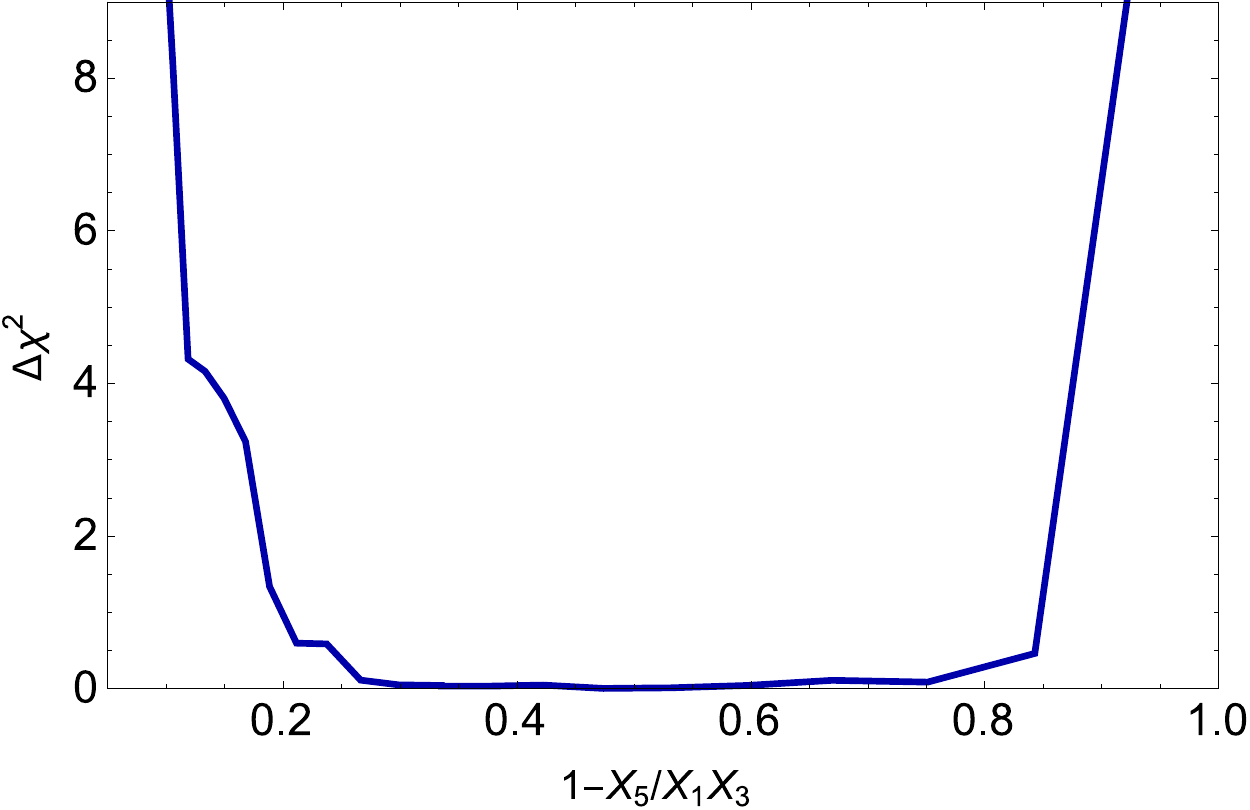}
   \caption{Left panel: $\chi^2$ profiles for the $X_i$ variables. Right panel: Diracness test  for a quasi-Dirac input point.}
  \label{fig:chi2profiles_XQD}
\end{figure}

\section{Conclusions}
\label{sec:conc}

We have studied the sensitivity of the DUNE and JUNO experiments to quasi-Dirac
neutrino oscillations. We have found that, within this scenario, the
determination of neutrino mixing angles becomes much more complicated,
if not impossible, even for next-generation experiments as DUNE or JUNO.
The main reason is that the new angles and phases are strongly
correlated to the corresponding ones in the active sector, leading to
very relevant degeneracies.

As a further comment, let us mention that many of the degeneracies
observed here could be broken by including a $\nu_{\tau}$ appearance
channel in the DUNE analysis.  This possibility has been recently
discussed in Ref.~\cite{deGouvea:2019ozk,Ghoshal:2019pab}.  If neutrinos are
quasi-Dirac particles, $\sum_{\beta}P_{\alpha\beta} < 1$, with
$\beta=\{e,\mu,\tau\}$.  Hence, a more precise observation of the
unitarity of neutrino oscillations including the $\nu_\tau$ channel
would be extremely helpful to test the quasi-Dirac neutrino
hypothesis, as well as other non-unitary neutrino
scenarios~\cite{Escrihuela:2015wra,Blennow:2016jkn}.

Despite the degeneracies affecting the angles, 
we have seen that we can define new observables which clearly allow to distinguish the standard
oscillation case from the quasi-Dirac neutrino scenario. While most of the
parameter space for these observables is still allowed at present, we have shown
that DUNE and JUNO can considerably improve the current bounds on these
quantities. We have also seen that, if quasi-Dirac neutrino
oscillations are real, the new generation of  experiments will have the potential to discover
quasi-Dirac neutrinos, which would be a big breakthrough in particle physics.

\section*{Acknowledgments}
We would like to thank Stefano Gariazzo and Renato Fonseca for
useful discussions.  Work supported by the Spanish grants
FPA2017-90566-REDC (Red Consolider MultiDark), FPA2017-85216-P and
SEV-2014-0398 (MINECO/AEI/FEDER, UE), as well as PROMETEO/2018/165
(Generalitat Valenciana). VDR acknowledges financial support by the
"Juan de la Cierva Incorporacion" program (IJCI-2016-27736) funded by
the Spanish MINECO. CAT is also supported by the FPI fellowship
BES-2015-073593. MT acknowledges financial support from MINECO through
the Ram\'{o}n y Cajal contract RYC-2013-12438.

\bibliographystyle{kp}

\begingroup\raggedright\endgroup

\end{document}